# Inherent fluctuation-mediated "equivalent force" drives directional motions of nanoscale asymmetric particles

## —— Surf-riding of asymmetric molecules in thermal fluctuations

Yusong Tu[+], Nan Sheng[+], Rongzheng Wan, Haiping Fang[*]

Division of Interfacial Water, Shanghai Institute of Applied Physics, Chinese Academy of Sciences, P.O. Box 800-204, Shanghai 201800, China

[+] Equal contribution

[*] Corresponding author (email: fanghaiping@sinap.ac.cn)

**Abstract**

Using a simple theoretical model of a nanoscale asymmetric particle/molecule with asymmetric structure or/and asymmetric charge distribution, here using a charge dipole as an example, we show that there is unidirectional transportation mediated by non-white fluctuations if the asymmetric orientation of the particle/molecule is constrained. This indicates the existence of an inherent "equivalent force", which drives the particle/molecule itself along the orientation of the asymmetric particle in the environment of fluctuations. In practical systems, "equivalent force" also exist in the asymmetric molecules, such as water and ethanol, at the ambient condition since thermal fluctuations are not white anymore at nanoscale [Wan, R., J. Hu, and H. Fang, *Sci. China Phys. Mech. Astron.* 2012, 55, 751]. Molecular dynamic simulations show that there is unidirectional transportation of an ultrathin water layer on solid surface at room temperature when the orientations of water molecules have a preference. The finding will play an essential role in the understanding of the world from a molecular view and the developing of novel technology for various nanoscale and bulk applications, such as chemical separation, water treatment, sensing and drug delivery.

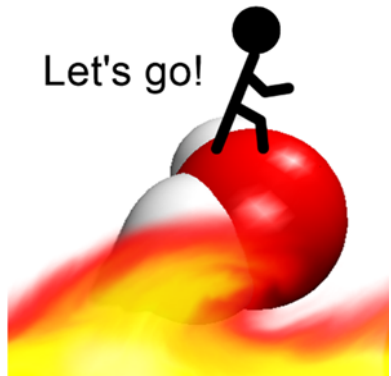

**Introduction**

Nanoscale systems usually exhibit behaviors different from their macroscopic counterpart. Up to now, most of studies on the nanoscale systems focus on effects of size and shape (1-4), and quantum effects (5-7). On contrast, much less concerns have been paid on the impacts from distinguished behaviors of environmental thermal noises at the nanoscale. Recently, we show that the thermal noise cannot be simply treated as white noise, i.e., the auto-correlation time of the thermal noise is comparable to the correlation lengths in nanoscale systems (8). Due to the finite time for the orientation regulation, asymmetric nanoparticles present spontaneous asymmetric diffusions with ~10% more possibilities for the particle moving along the initial orientation than moving, opposite, in the timescale of ~ 100 ps (9, 10), and asymmetric transportation could be induced by thermal noise at the nanoscale (8, 11).

Here, using a simple theoretical model of a charge dipole (with a typical charge asymmetry), we show that there is an inherent "equivalent force" along the dipole orientation for an asymmetrical nanoscale particle in the environment of fluctuations. If we constrain the orientation of the asymmetrical particle, there will be unidirectional transportation along the orientation. Using molecular dynamic simulations, we further show that this "equivalent force" results in a unidirectional flux of an interfacial water layer on the solid surface at room temperature if the orientations of the water molecules are constrained. The value of this unidirectional flux depends on the extent of the constraining on the orientations. The key to the existence of this "equivalent force" is the asymmetrical diffusion of nanoscale molecules/particles together with the non-white behavior of the fluctuations (the thermal fluctuations are not white at nanoscale).

**Simulation and results**

We construct a model with one dipole, namely, Main model. The dipole comprises two atoms, both of them have the mass, $m$, and the charge quantity, $q$, but with opposite signs. These two atoms are connected by a spring, as shown in Fig. 1(a-c). The atoms are indexed by $i$ ($i = 1, 2$ and the charge on the atoms are $q_1 = +q$ and $q_2 = -q$, respectively). The motion of the atom $i$ is governed by the Langevin equation

$$m\ddot{\boldsymbol{r}}_i = -\nabla U(\boldsymbol{r}_i) - \gamma m \dot{\boldsymbol{r}}_i + \boldsymbol{\eta}(t) \tag{1}$$

where $\boldsymbol{r}_i$ is the position of atom $i$, $\gamma$ is the damping constant, $k_B$ is the Boltzmann constant, $T$ is the temperature of the environment, $\boldsymbol{\eta}(t)$ represents the thermal fluctuations. $U(\boldsymbol{r}_i)$ is the interacting potential on atom $i$, which includes two parts, the harmonic potential between the atoms of dipole and an external potential $U(\boldsymbol{ex})$ to constrain the orientation of the dipole.

$$U(\boldsymbol{r}_i) = U(\boldsymbol{ex}) + 1/2\, k^{bond}(r - L)^2, \tag{2}$$

where $r = |\boldsymbol{r}_1 - \boldsymbol{r}_2|$ is the distance between the two atoms, $k^{bond}$ is the spring constant and $L$ is the equilibrium length between the two atoms. The external potential $U(\boldsymbol{ex})$ is obtained by applying a uniform electric field ($\boldsymbol{E}$) so $U(\boldsymbol{ex}) = -q_i \boldsymbol{r}_i \cdot \boldsymbol{E}$. The thermal fluctuations $\boldsymbol{\eta}(t) = \sqrt{2\gamma k_B T m}\, \boldsymbol{R}(t)$ $\boldsymbol{R}(t)$ is composed by square-wave functions with time period $\tau$ and magnitude of Gaussian random distribution and zero-mean in each dimension. Thus, $\boldsymbol{R}(t)$ satisfies

$$\langle \boldsymbol{R}(t) \rangle = 0 \tag{3}$$

$$\frac{\langle \boldsymbol{R}(t) \cdot \boldsymbol{R}(t') \rangle}{\langle \boldsymbol{R}(t) \cdot \boldsymbol{R}(t) \rangle} = \begin{cases} 1 - |t - t'|/\tau & |t - t'| < \tau \\ 0 & |t - t'| \geq \tau \end{cases} \tag{4}$$

Considering the charge asymmetry of dipole, the damping constants should be associated with the dipole direction (10). Here, we assume the average damping constant $\gamma$ as 0.50 ps$^{-1}$, and the damping constant $\gamma_F = 0.51$ ps$^{-1}$ when the dipole moving forward along its dipole direction, $\gamma_B = 0.49$ ps$^{-1}$ when moving backward. (Detailed parameters settings are refereed to method parts).

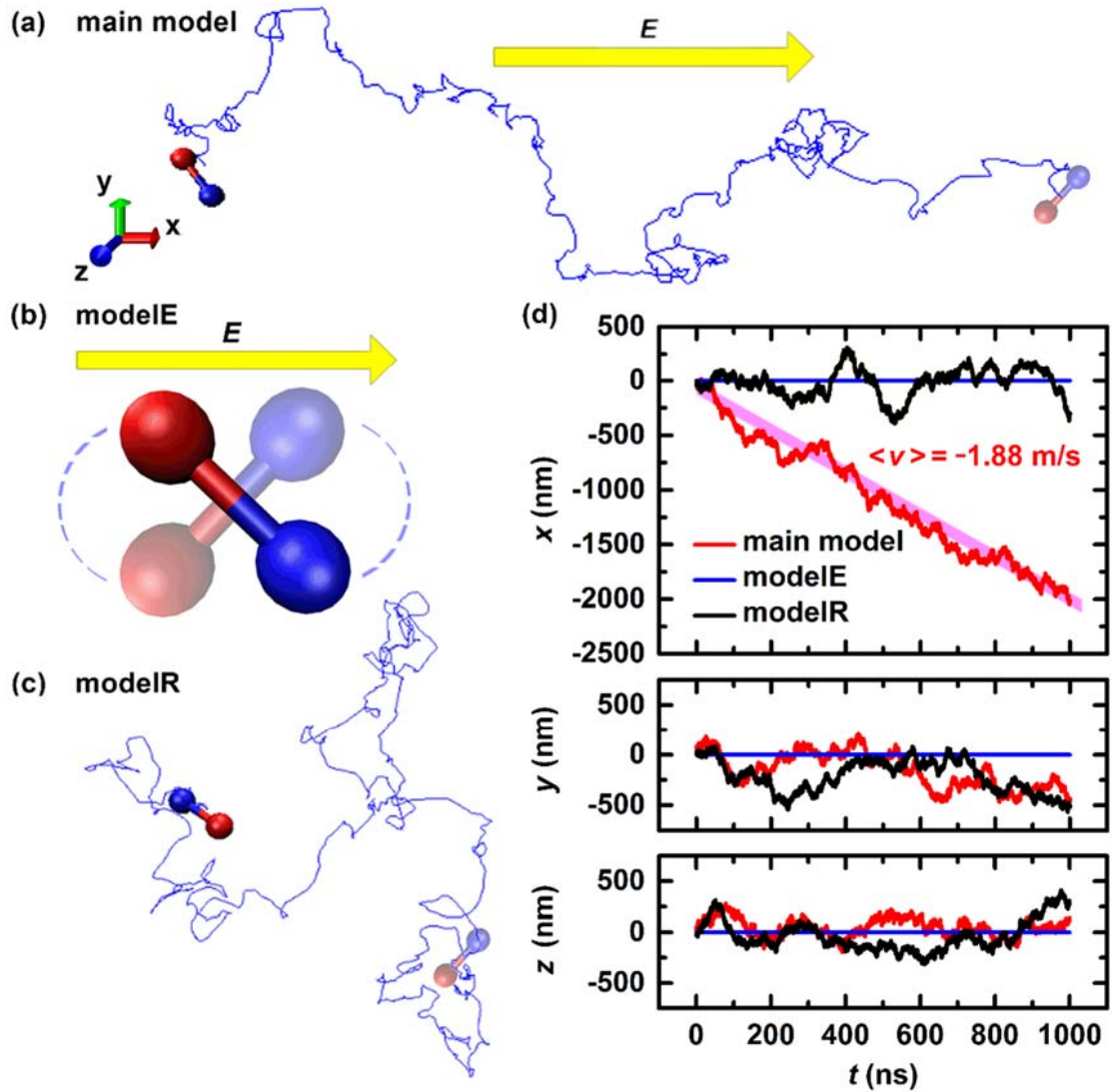

Figure 1. Motion of a charge dipole. (a-c) Typical trajectories for Main model, ModelE, and ModelR. The dipole is shown in spheres with a bond, with the negative charge in red, and positive, in blue. The initial positions are shown by shadowed dipoles. In Main model (a) there are both external static electric field and fluctuations; In ModelE (b) there is an external static electric field but no any fluctuation; in ModelR (c) there are fluctuations but no any external field; (d) Displacement of the center of mass (COM) of the dipole in Main model (red line), ModelE (blue line) and ModelR (black line) along x-direction with respect to the time. (e) Displacement of the COM of the dipole in Main model (red), ModelE (black) and ModelR (blue) along y direction (dash line) and z direction (dot line).

Figure 1(a) shows a typical trajectory for the main model, with E = 0.10 V/m and $\tau$ = 1.00 ps. We can see that the dipole has a directed diffusion along -x direction. The unidirectional transportation can be seen more clearly from the position of the center of mass (COM) of the dipole as shown in Fig. 1(d, e) (here we use the position of the COM to represent the dipole position). The displacement of the dipole along -*x* direction increases almost linearly, and the unidirectional transportation has an average velocity ⟨$v$⟩ of ~ -1.88 m/s. The displacements of dipole along the other two directions fluctuates with a very small mean values.

In order to show impacts of thermal noise and constraint of the orientations on dipole transportation, we further construct two new models, ModelE and ModelR, based on the Main model. Explicitly, if we set $\gamma = \gamma_F = \gamma_B = 0$ in the Main model (suggesting the environment without thermal fluctuation), we get ModelE; and we have ModeR if we set $E = 0$. Figure 1(b, c) shows the typical trajectories of two new models. In ModelE, the dipole oscillates around its equilibrium position and the amplitude is related to its initial positions; random diffusion of the dipole is observed in ModelR. The displacements of the dipole in Fig. 1(d, e) show that the COM of the dipole in ModelE is fixed to its initial position, and the COM of the dipole in ModelR fluctuates randomly and the mean displacements of dipole from the initial position shows non-directional behavior.

The key to the unidirectional transportation of the dipole in Main model is the ordering of the dipole orientation in the system. Figure 2 shows the normalized distribution of the dipole orientation $\theta$, where $\theta$ is the angle between the orientation of the dipole and the *x* direction. In ModelR, the flat distribution means the isotropic orientation of dipole, as expected, since the model is isotropic. In ModelE, the $\theta$ distribution depends on the initial position, at which there is a clear peak since the velocity of the dipole at this position is very small. Interestingly, in Main system, where both thermal fluctuation and external electric field exist, we can see the asymmetric distribution of the dipole orientation, where a peak turns up at $\theta = 0°$ and the value decreases monotonically as $\theta$ increases. Now we can say that the unidirectional transportation of the dipole happens only when there is a biased dipole orientation mediated by fluctuations.

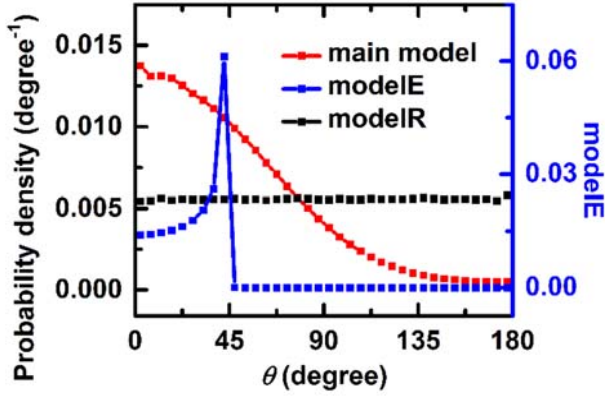

Figure 2. Distribution of the angle θ between the dipole orientation and *x* axis for Main model (red), ModelE (blue) and ModelR (black).

The explicit value of the velocity $\langle v \rangle$ depends on the preference of the dipole orientation in unidirectional transportation. Figure 3 shows that the average velocity $\langle v \rangle$ has a linear behavior with $\langle \cos\theta \rangle$ as

$$\langle v \rangle = C \cdot \langle \cos\theta \rangle , \tag{5}$$

where C is a constant. Here C = -4.11 nm ns$^{-1}$ from the best fitting.

We note that, if there is a constant force to pull a molecule along a fixed direction (for example, +*x*), there will be a final velocity along this direction. Thus, the existence of a constant average velocity $\langle v \rangle$ along + *x* direction indicates that there is an "equivalent force" *F* along +*x* direction. From Eq. (5), we have

$$F = \text{FUN} \cdot \langle \cos\theta \rangle, \tag{6}$$

where FUN is a constant. To evaluate the value of the constant FUN, we have applied a force *F* along +*x* in the modelR and find a linear relationship of $\langle v \rangle = C' \cdot F$, where the constant $C' = 125 \text{ nm}^2\text{ns}^{-1}\text{kJ}^{-1}\text{mol}$. Thus, we get FUN = -0.033 kJmol$^{-1}$nm$^{-1}$ in Main model (see detail in supporting information SI).

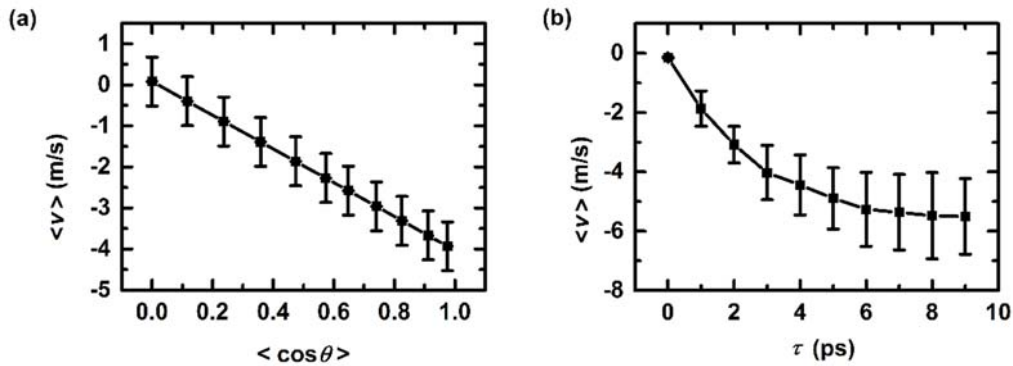

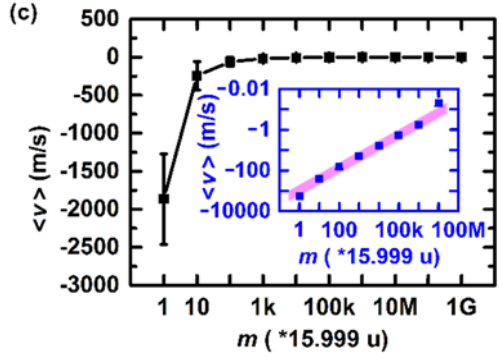

Figure 3 Average velocity ⟨v⟩ for unidirectional transportation along x axis. ⟨v⟩ with respect to (a) the average value of ⟨cosθ⟩, where $\tau$ =1.00 ps and m = 15.999 u, (b) the auto-correlation time $\tau$ where E = 0.1 V/nm and m=15.999 u, and (c) the mass m of the atoms, where E = 0.1 V/nm and $\tau$ =1.00 ps.

The constants C and FUN are associated with the structure of the dipole and the fluctuations. Figure 3(b) shows the value of ⟨v⟩ with respect to the characteristic time length $\tau$. We can see that ⟨v⟩ increases as $\tau$ increases from 1 fs and saturated from ~6 ps. We note that the velocity becomes quite small as $\tau$ = 1 fs, indicating the importance of the sufficiently long time length of the fluctuations in the formation of the unidirectional transportation at nanoscale. We have also study the behavior of the unidirectional transportation with respect to the mass of the atoms in the dipole as shown in Fig. 3(c), ⟨v⟩ decreases as power function with respect to the mass m.

Now we come to the practical systems. It is well recognized that many molecules, including water molecules, have asymmetrical structures. However, the thermal noise is usually taken as white noise so that the auto-correlation time is zero. According to the analysis above, there is no unidirectional transportation of those asymmetrical molecules under such thermal noise. It should be noted that, the thermal fluctuations are not white anymore at nanoscale, considering that the typical time duration between two collisions is on the order of 10 to 100 picoseconds (8), consistent with the statement by Magnasco that the timescale of thermal fluctuations is smaller than ~100 ps (12). Our MD simulations have further shown that the auto-correlation times are ~ 10 ps in the water at room temperature, independent of thermostat methods in the MD simulations (13). Thus, we expect that the "equivalent force" also exist in the asymmetric molecules in the ambient condition, and unidirectional transportation can also be observed in practical systems when the orientations of the asymmetric molecules constrained.

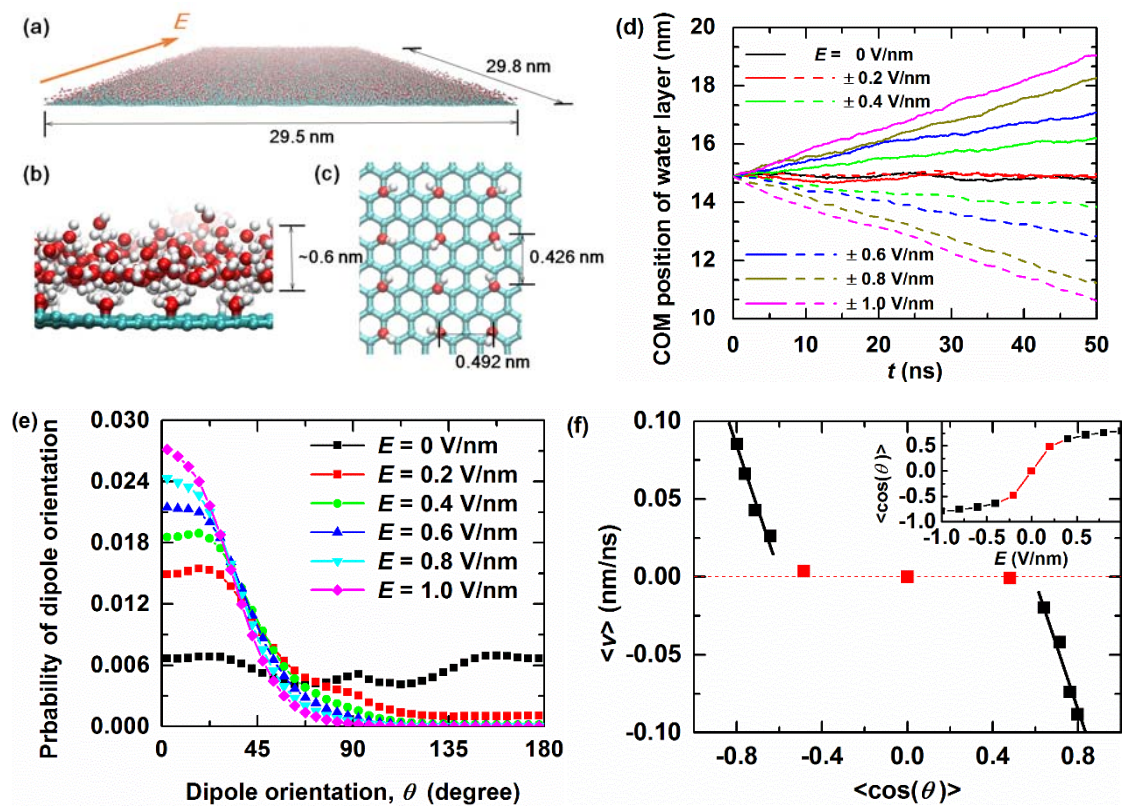

**Figure 4.** Unidirectional transportation of an interfacial water layer on the graphite-like surface under external electric field $E$ along $x$ direction. (a) Illustrating view and (b) side regional view of ultrathin interfacial water on a hydrophilic surface; (c) the geometry of the hydrophilic surface (top view). Water molecules are shown as spheres with oxygen in red and hydrogen in white; the surface carbon atoms, as well as bonds are shown in cyan; Hydroxyl groups are shown as bonded spheres with oxygen in red and hydrogen in white. (d) Displacement of the COM of the interfacial water layer along the x axis, in electric fields of magnitudes from -1 to +1 V/nm with an interval of 0.2 V/nm. (e) Probability of orientation angle $\theta$ of water molecular dipoles. (f) Dependence of $\langle v \rangle$ on $\langle \cos\theta \rangle$ in electric fields with various $E$ magnitudes, and the inset show the relationship between $\langle \cos\theta \rangle$ and $E$ magnitudes.

Here, as an example, we consider the behavior of an ultrathin interfacial water layer on solid surfaces. The surface used here is hydrophilic, which is constructed by 29.5 nm × 29.8 nm graphite-like lattice planted by hydroxyl group at every two carbon rings as shown in Fig 4(a-c). Initially, 10296 water molecules are placed on the surface and spread over the whole surface with a thickness of 2-3 layers of water molecules after 80 ns MD simulation.

Then we applied an electric field of $E$ along $x$ axis on the surface (the $x$ axis is defined along the armchair direction of graphite-like surface) to constrain the orientations of the water, and performed MD simulations for 50 ns for each magnitude of $E$. Figure 4(e) displays the normalized probability distributions of dipole orientations ($\theta$) of the ultrathin interfacial water layer. We can see that, for E ≥ 0.2 V/nm, there is a clear peak at $\theta$ close to 0º and the probability almost decreases monotonically as $\theta$ increases and reach a very small value as $\theta$ > 120º. The larger the magnitude of E, the higher the peak at $\theta$ = 0º. This indicates that water molecules always has the orientation preference along the direction of the electric fields; the stronger the electric field, the more significant the orientation preference. We note that similar behavior can be observed, when we applied the electric fields in the opposite direction and the water molecules have the orientation preference along the opposite direction. In the inset of Fig. 4(f), we show the average cosine values $\langle \cos\theta \rangle$ with respect to the various magnitudes of the applied electric fields. It is clear that $|\langle \cos\theta \rangle|$ increases monotonically and approaches to 1 as the absolute value of $E$ increases.

Figure 4(d) presents the COM displacements of the ultrathin interfacial water along the direction of the electric field. Unidirectional transportations can be seen clearly when |E| > 0.2 V/nm. Moreover, the displacements of water COM increase almost linearly, indicating constant velocities of transportations.

As indicated in the Main model system above, we can also obtain the linear behavior of the average velocity $\langle v \rangle$ with respect to $\langle \cos\theta \rangle$, as shown in Fig. 4(f). Explicitly, a very well linear fit can be achieved for the part of $|\langle \cos\theta \rangle| > 0.62$, as
$$\langle v \rangle = C \cdot (\langle \cos\theta \rangle \pm \cos\theta_0). \tag{7}$$
where C = -0.41 nm/ns, and $\cos\theta_0$ ~0.59, $\pm$ corresponds to the directions of electric fields along $\pm x$ axis. We think that the threshold $\cos\theta_0$ comes from the adsorption exerted on interfacial water molecules by the solid surface. Therefore biased orientations of water dipoles can take shape of an "equivalent force" $F$ on the surface.

In order to explicitly measure the "equivalent force", we removed the external electric fields so that there is no orientation preference of the water, but apply an external force to each water molecules along +x direction. We performed MD simulations for this new system and found a linear relationship of $\langle v \rangle = C' \cdot F$ with $C'$~2.04nm$^2$ns$^{-1}$kJ$^{-1}$mol (see Fig. S2). Thus, FUN ~ -0.2 kJmol$^{-1}$nm$^{-1}$ (see the detail in supporting material S2).

The existence of unidirectional transportation is universal to all directions along which the electric field is applied to enable the orientations of the water dipoles to have preference. However, both the fluctuations of the COM and the COM velocity of the COM with respect to $|\langle cos\theta \rangle|$ will be different, because of different details of the surfaces. Figure S2 show the displacement of the COM of the interfacial water layer along the zigzag direction of graphite-like surface, without any external electric field and with an external electric field along this direction. We can see that there are large fluctuations, especially for the case without any external field (E = 0). The velocity for E = 1V/nm becomes 0.03nm/ns. We think that these differences come from different water-surface interactions along different directions.

**Conclusion**

We use a simple theoretical model to demonstrate the existence of an inherent "equivalent force" along the dipole orientation of asymmetrical nanoscale molecules/particles mediated by fluctuations. The "equivalent force" can be embodied in unidirectional motion by constraining the orientations of the nanoscale molecule/particle; and their speeds depend on the extent of the orientation constraining. Herein, the auto-correlation time length in fluctuations plays the key role, i.e., the "equivalent force" disappears when the auto-correlation time becomes zero. In an example of practical systems, we show unidirectional transportation of an ultrathin water layer on solid surface when the orientations of water dipoles are constrained by an external uniform electric field at room temperature. As the simple model, the magnitude of the "equivalent force" is in proportion to the ordering degree of water molecular dipoles (the average cosine of orientation angles of water dipoles), and, the non-white behavior of the thermal noise (the autocorrelation time is ~ 10 ps for thermal noise (8)) is the key for the existence of this "equivalent force". It should be noted that the directed moving of an asymmetrical particle along its orientation direction mediated by the thermal fluctuations just like surf-riding of this particle in thermal fluctuations. This work explains the existence of the unidirectional transportation across the nanotube in Ref. (14, 15), and the results for the single-file water in Ref. (16, 17). The findings will play an essential role in the understanding of the world from a molecular view and the developing of novel technology for various nanoscale and bulk applications, such as chemical separation, water treatment, sensing and drug delivery.

## System and Method

**Simulations on the model system**. We carried out Langevin dynamic simulation for the transportation of one single dipole with asymmetric damping coefficients along dipole orientation, and the dipole orientation was constrained by electric fields. The temperature is set as $T = 300$ K. Velocity Verlet integration is used for 1 μs simulation, and a time step of 1 fs is used. The parameters on the dipole are set as below: charge quantity, $q = 0.840$ e; the dipole displacement, $L = 0.150$ nm. Considering that the damping constant is associated with the orientation of the molecule (10), the asymmetric damping constants can be introduced into systems, i.e., $\gamma_F = 0.51$ ps$^{-1}$ when moving forward along dipole direction, and $\gamma_B = 0.49$ ps$^{-1}$ when the dipole move backward. The average damping constant $\gamma$ is set as 0.50 ps$^{-1}$.

**Simulations on an interfacial water layer on the graphite-like surface.** The molecular dynamics simulation has been recognized as one of the most effective tools in the study of the dynamics of the systems on the nanoscale dimensions (14, 15, 18-34). We carried out molecular dynamics simulations for diffusion of interfacial water in a uniform static electric field, with GROMACS software package version 4.5.5 (35, 36). The system include 10296 SPC/E water molecules on a model graphite substrate surface with dimensional size 29.52 nm × 29.82 nm × 5.2 nm, where 4200 hydroxyl- (–OH) groups were "planted" at every two carbon rings to achieve a hydrophilic surface. A ceiling surface laying 4.5 nm higher than the substrate surface was employed to prevent water molecules from reaching the other side of substrate surface due to the periodic boundary condition. The static electric field is assigned along the *x* axis in systems, in parallel with the substrate surface. The carbon atoms of graphite surface are treated as uncharged Lennard-Jones particles with a cross-section of $\sigma_{CC} = 0.3581$ nm, and a potential well depth of $\varepsilon_{CC} = 0.2774$ kJ mol$^{-1}$; the parameters of Hydroxyl-group are taken from the GROMOS96 version 53a6 force-field for serine, including the charges, $q_O$=-0.674 e, $q_H$= 0.408 e and $q_C$= 0.266 e. All carbon atoms of graphite were position-restrained, but hydroxyl- (–OH) groups is applied with standard angle potential to keep C-C-O angle as 90° and C-O-H angle, 109.5°. The rotations of Hydrogen atoms around C-O bonds are in principle possible but are effectively quenched due to hydrogen-bonding interaction with interfacial water.

The NVT ensemble is used, with temperature controlled at 300 K by the velocity-rescale thermostat with a coupling coefficient of $\tau_T$ = 0.1 ps (37). The particle-mesh Ewald method was used to calculate the long-range electrostatic interactions (38), whereas the vdW interactions were treated with smooth cutoff at a distance of 12 Å. A time step of 2.0 fs was used, and data were collected every 2 ps.

**Acknowledgements**

We thank Prof. Bailin Hao for the important remark "Biomolecules surfing in thermal fluctuations" and very valuable discussions, and Prof. Jun Hu, Dr. Xiaoling Lei and Mr. Jian Liu for helpful discussions. This work was supported by the National Natural Science Foundation of China (Grant Nos. 11105088, 11175230 and 11290164), the Supercomputer Center for the Chinese Academy of Sciences and Shanghai Supercomputer Center.

# supporting materials

## S1. Calculation of FUN in Main model

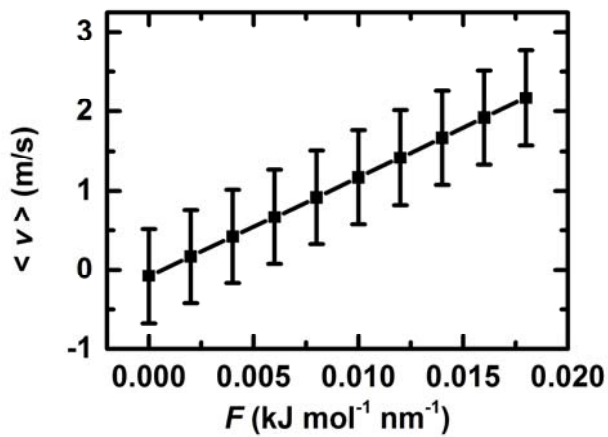

**Figure S1.** Average velocity $\langle v \rangle$ for unidirectional transportation along $x$ axis with respect to the "equivalent force" $F$.

From Fig. S1, the average velocity $\langle v \rangle$ has a linear behavior with $F$ as
$$\langle v \rangle = C' \cdot F$$
where $C'$ is the constant, $C' = 125 \text{ nm}^2\text{ns}^{-1}\text{kJ}^{-1}\text{mol}$ from the best fitting. Together with linear equation $\langle v \rangle = C \cdot \langle \cos \theta \rangle$, $C = -4.1$ nm ns$^{-1}$ from the manuscript, we can get
$$F = \frac{C}{C'} \langle \cos \theta \rangle$$
According to the equation of $F = \text{FUN} \cdot \langle \cos \theta \rangle$, we can get
$$\text{FUN} = \frac{C}{C'} = -0.033 \text{ kJmol}^{-1}\text{nm}^{-1}$$

## S1. Calculation of FUN in the superficial water on surfaces

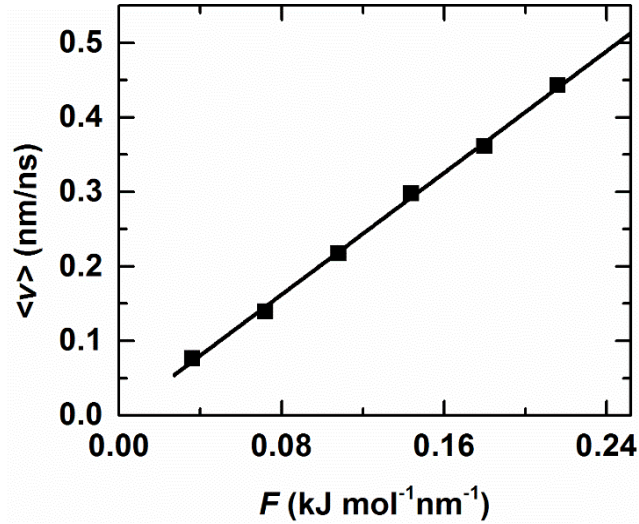

**Figure S2.** Average velocity $\langle v \rangle$ for unidirectional transportation along $x$ axis with respect to the "equivalent force" $F$.

We removed the external electric fields so that there is no orientation preference of the water, but apply an external force to each water molecules along +x direction. From Fig. S2, the average velocity $\langle v \rangle$ has a linear behavior with $F$ as

$$\langle v \rangle = C' \cdot F$$

where $C'$ is the constant, $C' \sim 2.04 \text{nm}^2\text{ns}^{-1}\text{kJ}^{-1}\text{mol}$ from the best fitting. Together with linear equation $\langle v \rangle = C \cdot \langle \cos\theta \rangle$, $C$ = -0.41 nm/ns from the manuscript, and thus we can get

$$F = \frac{C}{C'} \langle \cos\theta \rangle$$

According to the equation of $F = \text{FUN} \cdot \langle \cos\theta \rangle$, we can get

$$\text{FUN} = \frac{C}{C'} = -0.2 \text{ kJmol}^{-1}\text{nm}^{-1}$$

**S1. Unidirectional transportation of an interfacial water layer along the armchair direction of graphite-like surface where an external electric field is applied**

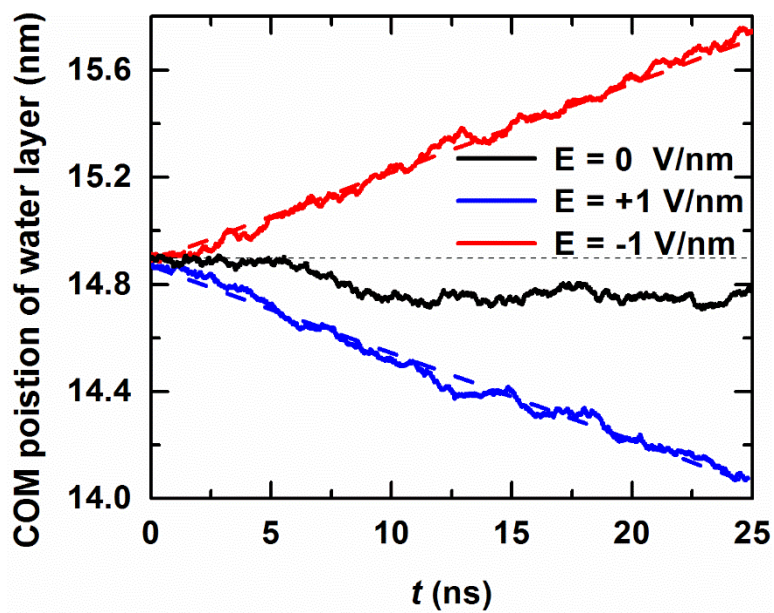

**Figure S3.** Displacement of the COM of the interfacial water layer along the zigzag direction of graphite-like surface, where an electric field is applied with magnitudes of E = 0, -1 and +1 V/nm.

Figure S3 shows the COM displacements of the interfacial water along the armchair direction of graphite-like surface. We can see that, without any external field (E=0), the COM of ultrathin interfacial water has larger drifts along the armchair direction. We have further performed MD simulation to the system with an electric field of $E$ along the armchair direction of graphite-like surface. The COM displacements of water increase almost linearly, indicating constant velocities of transportations. By comparison with results along the zigzag direction, we can see larger fluctuations in the directional transportations too, and the velocity for $|E|$ = 1V/m becomes 0.03nm/ns.